**MetaboTools: A comprehensive toolbox for analysis of genome-scale metabolic models**

Maike K. Aurich[1], Ronan M.T. Fleming[1], and Ines Thiele[1*]
[1]Luxembourg Centre for Systems Biomedicine, University of Luxembourg, Esch-sur-Alzette, Luxembourg.

[*]Corresponding author: Ines Thiele, Luxembourg Centre for Systems Biomedicine, University of Luxembourg, 7, avenue des Hauts-Fourneaux, L-4362 Esch-sur-Alzette, E-mail: ines.thiele@uni.lu.

**Abstract**
Metabolomic data sets provide a direct read-out of cellular phenotypes and are increasingly generated to study biological questions. Our previous work revealed the potential of analyzing extracellular metabolomic data in the context of the metabolic model using constraint-based modeling. Through this work, which consists of a protocol, a toolbox, and tutorials of two use cases, we make our methods available to the broader scientific community. The protocol describes, in a step-wise manner, the workflow of data integration and computational analysis. The MetaboTools comprise the Matlab code required to complete the workflow described in the protocol. Tutorials explain the computational steps for integration of two different data sets and demonstrate a comprehensive set of methods for the computational analysis of metabolic models and stratification thereof into different phenotypes. The presented workflow supports integrative analysis of multiple omics data sets. Importantly, all analysis tools can be applied to metabolic models without performing the entire workflow. Taken together, this protocol constitutes a comprehensive guide to the intra-model analysis of extracellular metabolomic data and a resource offering a broad set of computational analysis tools for a wide biomedical and non-biomedical research community.

**1. Introduction**
Omics data are used to determine comprehensively qualitatively or quantitatively cellular components and how they change across different conditions [1-4]. Of those, metabolomic data are the closest to an observed phenotype [5-7]. Consequently, metabolomics is becoming an indispensable analytical method for many biological disciplines, including microbiology, plant sciences, biotechnology, and biomedicine [8-10].

However, the analysis and interpretation of metabolomics data is still in its infancy, limiting the interpretation to few metabolic pathways rather than providing a comprehensive understanding of the underlying mechanistic basis [1]. At the same time, computational modeling methods, such as constraint-based metabolic modeling [11], are becoming increasingly popular for the interpretation of omics data [12] and for the generation of experimentally testable hypotheses [13-15].

Metabolomics data can be obtained from quenched cells (intracellular metabolome) or from the spent medium of cells in culture (extracellular metabolomic data). Extracellular metabolomic data sets are generated from cultivations of cell lines in order to metabolically characterize them under different experimental conditions (e.g., drug treatment and hypoxia) [16-18]. The extracellular metabolome captures metabolite consumption and byproduct release, e.g., lactate secretion by cancer cells, which can be interpreted as readout of the intracellular pathway use, such as aerobic glycolysis in cancer cells [7]. Whereas the connection between lactate and glycolytic flux is quite straightforward, the integration of entire uptake and secretion profiles can lead to novel insights of complex interactions between multiple pathways, which may be difficult to unveil manually. Previously, the intra-model analysis of extracellular metabolomic data of two T-cell lines let to the prediction of phenotypic properties of the cell lines, namely differences in flux through the TCA cycle and the electron transport chain, which were subsequently experimentally supported [7]. Computational analysis enables the prediction of the intracellular pathway activity that explain the measured metabolite uptake and secretion pattern and differences thereof between, e.g., cell lines or environmental conditions, and can lead to experimentally testable hypothesis.

The advantage of using extracellular metabolomic data lies in the accessibility of the medium, which saves time in sample preparation and allows for repeated measurements from the same cells. Moreover, concentration changes in the spent medium resulting from uptake and secretion by the cells can be converted into fluxes and used as constraints on the exchange reactions.

Extracellular metabolomic data have been extensively used in metabolic modeling to define cell and tissue specific exchange pattern [4, 19-23]; however, the process has never been addressed by a protocol explicitly. In comparison to the integration of transcriptomics data with metabolic models [24, 25], tools for integration of metabolomics data have not yet been made assessable to a broader research community. Some methods for the generation of contextualized metabolic submodels allow consideration of intracellular metabolomics data and the presence of reactions producing a set of detected metabolites is ensured [26-28]. However, the downstream analysis of contextualized metabolic submodels and their predicted metabolic phenotypes is not supported by these methods. The constraint-based modeling and analysis (COBRA) toolbox [29] provides an extensive set of functions for the computational analysis of metabolic models and a tutorial is available assisting in the interpretation of model predictions [30]. Despite the presence of these resources, a step-by-step guide, which captures the entire workflow of data analysis and phenotype prediction, is currently not available. With this protocol, the MetaboTools, and the accompanied tutorials, we close this gap and make a comprehensive set of methods available to the broader research community (Figure 1).

This work was developed and tested with our recent work [7], where we integrated extracellular metabolomic data of two T-cell lines with a human metabolic model to characterize the emergent phenotypic properties *in silico*. Furthermore, we recently mapped published metabolomic data from the NCI-60 cell lines [31] onto a human metabolic model and developed a suite of computational analysis tools that can be used to predict distinct metabolic features, e.g., the use of distinct pathways for energy production by the cancer cell lines [32]. The protocol discusses important considerations at individual steps to ensure successfully completion of the workflow (Figure 1). The tips and stipulations are derived from the aforementioned publications but also from our experience with curation and expansion of the human metabolic model using metabolomic data [33-35].

*This protocol covers:* This protocol provides support for the integration of metabolomics data into the network context and the generation of contextualized models. These contextualized models comprise a subset of the metabolic model and are primed to the prediction of the intracellular pathways, which may give rise to differences in the uptake and secretion profile of different cells or under different environmental conditions, using the minExCard method [32]. Additionally, this work provides tools for the analysis of any metabolic model.

The tutorials exemplify the workflows for intra-model analysis for (1) quantitative and (2) semi-quantitative extracellular metabolomic data (Figure 1, see also supplemental tutorial I and tutorial II), and make it facilitate the reproduction of our previous work [7, 32]. Importantly, the tutorials demonstrate the downstream analysis of the generated models. Additionally, we provide the data that are needed for the different integration steps (Table 1) and we discuss traits of metabolomic data sets to provide for a basis for their successful integration into the metabolic model (Table S1).

*This protocol does not cover:* The protocol provides limited discussion on the integration of other omics data. The compatibility of MetaboTools with the COBRA toolbox [29] allows the user to apply additionally the transcriptomic data analysis tools provided in the COBRA toolbox [29]. The protocol does not describe the intra-model analysis of intracellular metabolomic, untargeted metabolomic, or isotope labeling data. Furthermore, this protocol does not cover any steps concerning cell culture, mass spectrometry, data processing, and metabolite annotations. We ask the reader to refer to literature and dedicated tools from the respective fields [36-38].

## 2. Experimental design

**The protocol is divided into three stages.** The first stage provides the basis for the integration of extracellular metabolomic data, i.e., it ensures that a maximal number of metabolites can be integrated with a model. The second stage discusses the application of constraints and the generation of contextualized models. The third stage discusses the quality control of the contextualized models, the computational analysis tools provided by MetaboTools, and finally the validation of the model predictions. Several iterations of the steps in the second and the third stage may be needed to generate high-quality contextualized models and to obtain biologically-plausible model predictions (Table 2). Throughout the text, functions are written in italic. Input and output variables are indicated by asterisks (*…*). Matlab code is indicated through >>. Flux units are commonly reported in the Unit mmol/$g_{dry\ weight}$/hr (U); however, the unit can be varied depending on data [32].

**Stage 1 - Preparation of extracellular metabolomics data and models.**

**Associate metabolite IDs of the data with the metabolic model (Step 1):** As a first step, the names of detected metabolites need to be associated with the metabolite abbreviations in the model (Figure 2A, see supplemental material for an introduction to the model structure). Different standards exist to report metabolite identity and the human genome-scale reconstruction contains annotations for multiple identifiers, e.g., KEGG [39], ChEBI [40], HMDB [41], because none of the databases covers all its metabolites [35]. Association of a detected metabolite with a wrong counterpart in the model can lead to irrelevant predictions and conclusions. Moreover, new reactions might be introduced into the model as a consequence of the association and reused by researchers in or outside the working group, often without questioning why these reactions were added (Figure 2C). Hence, manual association of the metabolite identifiers is the method of choice, even though tools have been developed to facilitate the matching (e.g., [42]). The association is simplest based on, e.g., KEGG or HMDB metabolite identifiers. When matching metabolite names, one should consider synonyms and alternative naming conventions, e.g., palmitic acid is the traditional name of hexadecanoic acid (common name). Metabolite formulae are non-unique, and thus can only be used as additional clue to match metabolites, e.g., glucose and fructose have both $C_6H_{12}O_6$ as formula.

The step results in one group of metabolites successfully associated with model metabolite IDs and a second group of metabolites that does not yet exist in the metabolic model. The addition of novel anabolic or catabolic pathways to include the latter group of metabolites into the model can be time-consuming and requires extensive work as well as testing of the model functionality. The necessary steps have been described in detail elsewhere [43]. Thus, only the steps needed to prepare the integration of associated metabolites will be considered in this protocol.

**Can metabolites be transported into and out of the cell? (Step 2):** Although a metabolite is present in the model, this does not mean that it can be transported between the intracellular [c] and the extracellular [e] compartment (Figure 3A). Hence, it needs to be confirmed that transport reactions exist for all associated metabolites (Figure 2B). Transport reactions (e.g., ATP-dependent transport) can be irreversible, and transport reactions might need to be added that allow secretion or uptake of a metabolite. One exchange reaction needs to exist in the model for every associated metabolite (Figure 2B), since these (artificial) reactions mediate the supply or removal of metabolites to and from the extracellular environment of the model. These exchange reactions are used for the integration of extracellular metabolomic data into the model (Figure 3B).

**Identify missing metabolite transporter (Step 3):** Metabolomic data sets often contain metabolites, for which no transport and exchange reactions exist in the models, since i) high-throughput techniques detect more and more comprehensively metabolites in extracellular environments (e.g., body fluids or spent medium), ii) these metabolites were outside the scope of previous reconstruction efforts and applications [34], iii) their existence was unknown, or iv) their metabolism was unknown (Figure 2B, 3A).

The correct representation of transport mechanisms is important to accurately simulate cellular metabolism; thus, the metabolic reconstructions are continuously extended [7, 34, 35, 44, 45]. Hence, before getting started with this step, ensure that the most recent reconstruction version is obtained (e.g., the human metabolic reconstruction is downloadable from https://vmh.uni.lu/).

If extracellular transport reactions are missing for certain metabolites, they need to be identified from the literature. The identification of transport systems can take considerable time, which varies, depending on the number of metabolites and the extent to which the corresponding transport systems have been characterized [34]. Diffusion reactions should only be added if the transport system is unknown or diffusion of the metabolite has been reported. The model can use diffusion reactions to transport the metabolite "for free". As a consequence, energy and material costs of the metabolite transport will be underestimated in simulations. Thus, the exact transport mechanisms, all alternate and co-substrates, transport proteins, isozymes and ratio of subunits in protein complexes and their encoding genes, need to be identified. Correct gene-protein-reaction associations (see Box 1) are particularly important for integrating transcriptomic or proteomic data as well as for investigating the effect of genetic alternation in metabolite transporters. After the literature has been mined thoroughly, the new mass- and charge-balanced transport reactions have to be formulated [43] and the transport reactions need to be added to the model (Figure 2C). References to the primary studies, on which the addition of the transporter is based, should be documented [43].

**Addition of the transport and exchange reactions (Step 4):** Transport reaction should be added in a quality controlled manner to avoid typos in the metabolite abbreviations or whitespaces in the reaction formula. Typos easily go unnoticed until much later, since the functions to add reactions in the COBRA toolbox also automatically add new metabolites. Hence, reactions should be added in a quality controlled manner using rBioNet [46]. The addition of reactions (and metabolites) to the source SBML file is discouraged as they circumvent any quality-assurance and quality-control measures, and as such, are often a source of errors in the resulting model.

**Stage 2 - Addition of constraints and model generation**

**Define basic constraints (Step 5):** Metabolic networks are often distributed as reconstructions and not as condition-specific models. Hence, exchange and internal reactions are unconstrained, i.e., they have "infinite bounds" (Figure 4, Box 1, supplemental material). Alternatively, models may be distributed that mimick with their constraints a particular environmental or genetic condition, e.g., enabling growth on minimal medium under anoxic condition. Hence, the existing constraints of a reconstruction or model may not be a useful starting point for the one's own application. Whereas the directionality of transport reactions and internal reactions is consistent with current biological and thermodynamical data, the definition of the extracellular medium, i.e., the constraints on the exchange reactions (Figure 4), depends on the cell type and condition one wants to simulate. The application of a context-specific set of constraints on the exchange reactions ensures, together with the network topology, that the metabolic model closely resembles the cell type and experimental condition one wants to investigate, e.g., oxygen needs to be restricted to investigate hypoxia. The exchanges are best defined based on the experimental condition that one aims to model, e.g., culture medium composition, substrate and oxygen uptake rates. In absence of matching experimental data, literature values may substitute missing information [7, 47], but it should be noted that they can vary substantially between experimental setups and hence, may affect the prediction accuracy of the condition-specific model.

The function *setMediumConstraints* allows the definition of the model constraints and provides options for various configurations. If the composition of the cell culture medium is defined, the metabolite concentrations can be converted into fluxes that define metabolite uptake in the model using the function *setMediumConstraints* (Figure 3B). For this, cell number, cell dry weight (supplemental material), and experiment duration need to be known. The rationale of the added constraints is to restrict the model's metabolite uptake flux to the amount that was available to one cell and per 1 hour of the experiment. The medium composition can thus be used to reproduce *in silico* the experimental, or cell-type specific, condition (Figure 3B).

The model requires certain inputs and outputs to have a non-zero value for an objective function. For example, for the production of biomass of a human cell, the uptake of essential amino acids, ions, and other compounds needs to be provided to the model to render the objective function feasible (i.e., non-zero; see Box 1). Essential uptake reactions can be identified, e.g., using flux variability analysis (see Box 1).

The function *setMediumConstraints* includes an option to change the infinite bound (which is often defined as -1000 U for reverse reaction flux and +1000 U for the forward reaction flux), if it is necessary to prevent that the model is artificially constrained by imposed "infinite" bounds (see Step 16B). If the growth rate, or doubling time, for the given experiment is available, it can be set as constraint on the biomass reaction using the function *setMediumConstraints* or *changeRxnBounds*.

**Integration of metabolomic data (Step 6-13):** Metabolomic data can be integrated with metabolic models by enforcing exchange rates to uptake or secrete in accordance with the experimental data (Figure 3B-C). To enforce the respective directionality, minimal flux values for uptake and secretion should be defined, e.g., by using information on the detection limit for each metabolite (Figure 4). The underlying rationale is that as the metabolite has been measured at a higher or lower level after a certain time and hence, the secretion/uptake had to be above the corresponding detection limit. The conversion of the limits of detection from ng/ml to mM can be done using the function *calculateLODs* and using the molecular weight of the metabolites (see supplemental material). Uptake and secretion profiles for each sample are generated from an input data matrix using the function *defineUptakeSecretionProfiles*. See supplemental tutorial I for an example of the input data matrix. The directionality of exchange is defined based on the change over time in the spent medium and with respect to the change in the controls to rule out effects of spontaneous metabolite degradation (Figure 5A-B). The direction of the exchange might change when the controls are taken into consideration because the signal might drop over time in these medium controls, whereas it seems to remain stable in the cell culture medium due to secretion of the cells. Such cases need to be looked at carefully and it might be worth to exclude those data points. Mass spectrometric measurements are not exact and generally associated with an error. The calculated change between control and sample is also highly sensitive to the SD. A calculated net change of 5% can be meaningless if the SD is 10%. Hence, caution should be taken when incorporating changes in metabolite abundance, when the change is below or close to the SD. The tutorials illustrate how the constraints can be adjusted to one's data set at the different parts of the workflow.

The uptake and secretion profiles are combined with the detection limits using the function *calculateQuantitativeDiffs* and integrated with the model using the function *setQualitativeConstraints*. As a consequence, the flux through an exchange reaction must lay between the smallest detectable and the highest possible flux in case of uptake (defined by the medium composition), and between the minimal detectable efflux and the 'infinite' flux value for secreted metabolites (Figure 4).

In an additional step, relative differences in metabolite uptake and secretion can be integrated to compare pairs of models (Figure 4, 5C). The relative differences are defined using the function *calculateQuantitativeDiffs* based on the comparison of change between the samples and with respect to the controls (slope ratio, Figure 5A). Subsequently, the quantitative differences are applied to the two models (Figure 3, 5A, C) using the function *setSemiQuantConstraints*.

If absolute concentrations have been obtained for at least two time points, they can be converted into fluxes using the function *conc2rate* and applied as bounds on the exchange reactions considering a user-defined error (Figure 4). Individual uptake and secretion profiles are produced from an input data matrix of flux values with samples (columns) and metabolites (rows) using the function *prepIntegrationQuant* (see tutorial II). Negative values will be interpreted as uptake and positive values are interpreted as secretion. Based on the input model and user-defined minimal and maximal values, the function *prepIntegrationQuant* tests whether the uptake and/or secretion of each individual exchange in the input data matrix is feasible, using flux balance analysis [30]. If a metabolite cannot be consumed or secreted by the model due to missing synthesis or degradation pathways, these metabolite exchanges will be removed automatically from the exchange profiles. If only the secretion is infeasible, the secretion value is eliminated from the profiles, whereas the uptake value of the same metabolite will be kept. The function *checkExchangeProfiles* can be used to generate statistics on the number and identity of uptake and secretions added per sample.

After individual uptake and secretion profiles have been generated for each sample, i.e., cell types or conditions, these can be integrated with the metabolic model using the function *setQuantConstraints*. The function *setQuantConstraints* offers the option to add or eliminate constraints, e.g., if quantities or combination of constraints render the model infeasible. The function also allows the user to specify a lower bound for the objective function, which ensures that the output model is able to grow or to perform a specified metabolic task (e.g., lactate production), while the upper bound remains unconstrained. Which objective function is chosen should be carefully decided based on the experimental context (see [48] for discussion on use of different biomass objective functions for human cells). The output of *setQuantConstraints* is a contextualized submodel for each sample. The integration of quantitative extracellular metabolomics data can be performed for large sample collections [32].

**Generation of contextualized metabolic models (Step 14)**: The function *setQuantConstraints* automatically generates a contextualized model for each sample by calling the function *generateCompactExchModel*. These contextualized models are based on a minimization of the cardinality of exchange reactions of the constrained model, which means that the number of exchanges, in addition to those defined by constraints (e.g., metabolomic data), is minimized in the resulting model [32]. As the function *generateCompactExchModel* relies on an approximation, which may not be minimal [32], the minimization is repeated until the number of added exchange reactions cannot be further reduced. Additional exchanges may be required as untargeted metabolomics methods can still not measure and identify the entire metabolome, despite continuous improvements in the field [49]. In contrast, targeted metabolomics methods measure only a defined subset of a given metabolome, and thus, more metabolites may be exchanged with the environment by the cell. Our computational approach was set up to deal with these technical limitations by adding the required minimal set of metabolites that would together with the defined uptake and secretion profile, explain the measured differences in the cell phenotypes [32].

After the minimal set of required exchanges is defined, blocked (i.e., flux inconsistent) reactions are identified [26] and a flux-consistent submodel ('pruned') is extracted. The pruned model contains the predicted minimal set of exchange reactions (including constraints based on the quantitative data) as well as an active set of internal reactions. The constrained pruned and the constrained unpruned model are returned from the function *setQuantConstraints* as a structure variable *ResultsAllCellLines*. Additionally, an overview table is returned (*OverViewResults*), which summarizes the numbers of reactions, metabolites, and genes for each contextualized model of the sample set. The information in these variables are extended by the downstream analysis functions (see Step 17).

Note that the function *generateCompactExchModel* only returns one of multiple alternative solutions. The relevance of the set of metabolite exchanges added to the model needs to be evaluated, e.g., by comparison with the carbon sources commonly used by the cell type or organism based on experimental data [32]. The function *generateCompactExchModel* can also be applied to generate a contextualized model after qualitative or semi-quantitative constraints have been applied.

The function *extractConditionSpecificModel* has previously been used to extract a flux consistent model in [7]. The 'minimal' set of exchanges identified through this function is also not unique and the number and chosen additional exchange reactions depended on the order, in which the tested exchange reactions were closed.

**Multi-omics integration (Step 15):** MetaboTools supports the integration of gene expression and proteomic data. The function *integrateGeneExpressionData* can be applied to qualitatively integrate transcriptomic or proteomic data along with the metabolomic data. The function disables reactions (i.e., it sets the lower and upper bound to 0) associated with the user-defined set of unexpressed genes. This stringent treatment may lead to an infeasible model, if any metabolite required for the objective cannot be produced anymore by the model. The manual assessment and curation of gene expression data integration has been previously described for a signaling model [33]. The same principles apply to a metabolic network. A multitude of other methods exists for the integration of transcriptomic data [24]. Whether to curate an infeasible model based on literature or to use a different approach for gene expression data integration (e.g., *createTissueSpecificModel* as implemented in the COBRA toolbox [29]) needs to be determined for each case.

Proteomic data can be integrated in a similar way as transcriptomic data. In this case, the protein identifiers need to be matched to the gene IDs in a model, whereby the same attention should be paid to the conversion of protein to gene IDs as for the association of metabolite IDs to the model metabolites. Subsequently, the set of gene IDs that is associated with the absent proteins constitutes the input for *integrateGeneExpressionData*. It is recommended to treat missing gene/protein data points as present rather than absent. The additional omics constraints can be applied either before integrating metabolomic data, after executing *setMediumConstraints,* or after generating the contextualized models (i.e., after executing either *setQuantConstraints* or *generateCompactExchModel*).

**Stage 3 - Model validation and prediction of phenotypes**

**Assessment of constraints and expectations (Step 16):** The workflow supported by MetaboTools allows rapid generation of contextualized models. Yet, during optimization, a minimal set of exchanges, consistent with the provided experimental data, will be chosen without consideration of the biology. Furthermore, the complexity and redundancy of the metabolic network allows the prediction of submodels and results that comply with the applied set of constraints and the biologically well informed network topology, but which may not be necessarily biologically meaningful. For instance, cellular processes, such as signaling and regulation, influence metabolism and might prevent possible functional states *in vivo* or *in vitro* but they are not included in the model. Thus, the applied constraints and predicted exchanges need to be manually evaluated against the biological knowledge before proceeding with further computational analysis. If falling below the error range or exceeding the infinite bounds, constraints might need to be scaled (see supplemental material). All added exchanges and directions of exchange should be controlled to identify missing or erroneous constraints. For example, in our previous study, the initially generated set of cancer models predicted consumption of superoxide and secretion of oxygen because the constraints had not been appropriately defined [32]. The example illustrates the importance of manual inspection and biological insight when generating condition-specific models from experimental data (Table 2). Table 3 lists questions to ask and data to consider during this initial validation of the contextualized models.

The ability of the contextualized models to grow at the same rate as in the experiment can also be used to validate the contextualized model [32]. The optimization of biomass can be applied to cells in the exponential growth phase, and optimizing for biomass assumes that the cells are thriving towards optimal biomass production (see for explanation on the biomass composition [43]). However, growth might not be the suitable when dealing with primary human cells, where alternative biomass reactions, e.g., for maintenance of cells, may be assumed [48]. If doubling times are available for the experiment, the function *setConstraintsOnBiomassReaction* can be used to incorporate constraints on the biomass reaction, while considering a user-defined measurement error (e.g., ±20%) to separate upper and lower bound. Even slight differences in the culture conditions may have an impact on the growth performance of the cells.

MetaboTools contains a set of functions to investigate different aspects of the metabolic model (Step 24A-F), particularly when generating multiple models from a large sample set. The choice of the analysis to be conducted depends on the biological question to be addressed. Prior knowledge of pathways to focus the analysis on will prevent getting overwhelmed by the amount of output data, and will help to identify inaccuracies in the model(s) before dedicating too much time on the analysis (Table 2). To facilitate the interpretation especially for large model sets, MetaboTools contains functions that organize the analysis output into statistics and digestible tables [32].

**Define essential genes across a set of contextualized models (Step 17A):** The function *analyzeSingleGeneDeletion* predicts and summarizes essential genes of one or multiple models (Figure 6). The function returns the number of essential genes per model and a table, which sorts all genes that appear in one or more models into the three categories: 'no effect', 'all KO', and 'partial effect' along with the frequency of each effect among the tested set of models. Based on this table, interesting essential genes can be identified, e.g., those affecting only a subset of models, or all models equally (Table 4). The results for the individual model are added to the *ResultsAllCellLines* structure.

**Identify essential reactions (Step 17B):** A gene being essential for a model means that one or more reactions (i.e., combinations of reactions), which are associated with this gene, need to carry flux in the model in order to satisfy the defined objective function. However, the reactions associated with a gene can be distributed across different pathways, and depending on the pathway, the reaction may be more or less interesting for a biological question. E.g., an antimetabolite designed to target a non-essential reaction is useless because it would not stop a cell from proliferating. Hence, the function *checkEffectRxnKO* identifys the subset of essential reactions that are associated with a specified set of (essential) genes.

**Investigate common and distinctive features of a set of models (Step 17C):** Common and distinctive features between condition specific models can be insightful, e.g., to build a generic tissue model from a set of context specific tissue models. The function *makeSummaryModels* generates both, a union model and an intersect model, which sum either the common or the superset of reaction, metabolites and genes of a set of contextualized models.

**Analyze flux splits (Step 17D):** Cells use different pathways to produce energy and other crucial metabolites [32]. The function *predictFluxSplits* predicts how a metabolite of interest (e.g., ATP) is produced or consumed by the different reactions in the models (Figure 7). The analysis is based on a flux vector generated by using parsimonious flux balance analysis [50].

The function *predictFluxSplits* can also be used to predict a 'metabolite yield', i.e., the sum of flux producing a metabolite (e.g., ATP) divided by the uptake flux of a user-defined carbon-source, e.g., glucose exchange. Because no additional constraints are applied for this analysis, the, e.g., ATP generated by the model will not only be produced from glucose but also other substrates available to the model depending on the constraints on the exchange reactions of the individual model. Nevertheless, the metabolite yield constitutes a valuable measure to stratify metabolic models [32].

The results generated by *predictFluxSplits* can be summarized using the function *sumFluxSplits*, which provides a table listing for each model the reaction that produces the highest amount of the metabolite (highest flux among the reactions producing/consuming the metabolite of interest). Another output table provides the flux values to allow a comparison of the metabolite producing reactions across the model set. Based on the different strategies of how models produce or consume different relevant metabolites, sets of models can be divided into phenotypes [32] .

**Sampling the solution space (Step 17E):** Sampling is a method to explore the metabolic phenotype of a contextualized model without relying on an objective function [47, 51]; and hence, it is particularly suitable for modeling of human cells and biomedical applications [48]. The function *performSampling* performs the sampling analysis, including a priori generation of warm-up points using the functions of the COBRA toolbox [29]. After a specified number of flux vectors (sampling points) has been collected from the solution space of the model, the result of the analysis is illustrated as the probability distribution of flux for individual reactions in the model [51, 52] (supplemental material). The function *summarizeSamplingResults* obtains statistics on the sampling points and for each reaction (mean, median, and minimal and maximum flux values determined through flux variability analysis). It also generates histograms of the probability distributions for a set of user-defined reactions (Figure 8). Distinct use of pathways or reactions (e.g., directionality) can be directly observed from the histograms [4, 7, 47].

**Predict response to environmental changes (Step 17F):** Contextualized models can react differently to perturbations of the existing constraints [32]. The function *performPPP* can be used to analyze the impact of stepwise variations of flux forced through two exchange reactions. The "behavior" of the model is equal to the model being able to satisfy the stated objective. In other words, a flux balance analysis is performed at each step of the analysis, which means after the constraints of one of the two exchange reactions have been modulated. The variations can be either uptake or secretion, and the direction can differ between the two exchange reactions. This analysis was used to test the models response to variations in oxygen uptake and lactate secretion [32]. The results of the analysis can be illustrated as a heat-map using the function *illustrate_ppp*. The size and the shape of the feasible solution space within the heat map was used to manually distinguish different phenotypes among the set of contextualized models [32].

**Validation of the model predictions (Step 18):** The predicted set of essential genes (*analyzeSingleGeneDeletion*), preferential pathway use (*predictFluxSplits*), and feasible uptake rates (*performPPP*), constitute hypotheses and need to be validated. The validation of model predictions is an indispensable part of the computational workflow. In many studies the results are experimentally validated to substantiate the computational results [7, 14, 22, 23, 53, 54]. Others relate the model predictions to the analysis of omics data sets from cell cultures or patients [55, 56]. In case model predictions cannot be experimentally validated, model predictions should be supported by literature evidences [57, 58]. As many results as possible should be validated, but at least all main results of the study. The validation of the model prediction completes the workflow for the integration of extracellular metabolomic data and phenotype prediction.

3. Materials

   3.1 Equipment
   
   Hardware
   
   o Personal computer
   
   Software
   o Matlab (Mathworks, Inc.)
   o COBRA Toolbox v2.0 & MetaboTools
   o fastFVA (http://wwwen.uni.lu/lcsb/research/mol_systems_physiology/software)
   o A linear programming solver. We recommend the industrial quality solver cplex (IBM Inc.) as it can be used under a free academic license. Note that *generateCompactExchModel* requires cplex (IBM Inc.) to be installed and called from Matlab.
   o Obtain model (e.g., download the most recent version of the human metabolic genome scale model and numerous gut microbe metabolic models from the virtual metabolic human database (VMH, https://vmh.uni.lu/).
   
   Knowledge
   o Knowledge on the use of COBRA toolbox
   o Knowledge on the use of matlab
   
   Additional training in COBRA can be acquired from
   o Detailed description on the installation of the COBRA toolbox v2.0, are provided in a previous protocol on the COBRA toolbox [29].
   o Consider the COBRA toolbox forum for help (https://groups.google.com/forum/#!forum/cobra-toolbox).
   o FBA primer supplemental tutorial [30]
   
   3.2 Equipment setup
   o Install Matlab
   o Download COBRA toolbox (incuding MetaboTools) from (https://github.com/opencobra/cobratoolbox) and add to Matlab path
   **Hint:** Specifically, follow procedure steps 1-4 of the COBRA toolbox protocol [29] to ensure a functional COBRA toolbox. Functions that are not further explained in this protocol are part of the COBRA toolbox and their use has been described in detail in the COBRA toolbox protocol [29].
   o Install solver
   o Obtain a metabolic model (e.g., https://vmh.uni.lu/)

3.3 Input data when integrating metabolomic data
- Mandatory input data Extracellular metabolomic data (minimum two time points to calculate fluxes, see tutorials for format requirements)
- For conversion of metabolomic data to fluxes: cell weight, cell concentration\ml, duration of the experiment
- Information about medium composition
- Serum in medium (yes/no).
- Detection limits of the measured molecules

3.4 Optional input data
- Growth curves/ doubling times (hrs.)
- Transcriptomic or proteomic data (A/P calls)
- Additional measurements (e.g., oxygen consumption, $CO_2$)

4. Procedure:
**Stage 1: Preparations for the integration of the data (Steps 1-5)**
  1) **Associate metabolite identifiers with metabolite IDs, Trouble Shooting**
     Manually relate the metabolite identifiers in the metabolomic data to the corresponding metabolite abbreviation in the model (Figure 2A). If no HMDB [41] or corresponding IDs exist, associate the metabolites based on the metabolite name using metabolite databases. Find the exact matches.

  2) **Identify metabolites that have no transport or exchange reaction, Trouble Shooting**
     Compile a list and check off metabolites that have transport reactions (e.g., gln_L[e] <=> gln_L[c]) and exchange reactions ('EX_gln_L(e)') in the model. This can be done manually using the VMH database (https://vmh.uni.lu/, Figure 2B) or the models reaction list:

     Alternatively use the function *findRxnsFromMets* to generate a list of reactions (*rxnList*) that are associated with the list of successfully associated metabolites (*metList*):

     \>\> [rxnList, rxnFormulaList] = findRxnsFromMets(model, metList)

     Check also the lower and upper bounds of the transport and exchange reactions to confirm that the directionality corresponds to the reaction formula.

     To print the reaction list use:
     \>\> a = printRxnFormula(model, rxnList,0,0,0,'',0);

     Note that the reversibility vector (*model.rev*) is not updated by all functions in the COBRA toolbox. Most functions rely on the definition of the lower and upper bounds rather than the reversibility vector. Use the following code to make the reversibility vector consistent with the lower and upper bounds:

     \>\> model.rev = zeros(length(model.rxns));

>> model.rev(find(model.lb<0)) = 1;

3) **Identify metabolite transporter**
Manually identify and define as detailed as possible the extracellular metabolite transport systems for all associated metabolites lacking extracellular transport reactions based on the relevant literature (Figure 2C) [43, 59]. Start, e.g., by typing "metabolite AND transport" into a search engine or the NCBI PubMed database (http://www.ncbi.nlm.nih.gov/pubmed/). Variations of the key words will provide you with a first impression on the amount of literature that exists on the transport systems of individual metabolites. Transporters vary between compartments. Hence, specifically identify extracellular transport systems. Check for organism, tissue- and cell-type specific metabolite transport mechanisms. Does the literature describe or indicate active or passive transport (primary transport, secondary transport, or simple diffusion)? Identify additional substrates and co-substrates (such as ions, energy currency, or other metabolites), as well as the direction of transport (antiport versus symport). What is the reaction stoichiometry of the transported compounds? Identify the genes that encode the transport proteins. Consider the existence of isoforms and multiple subunits of protein complexes. Keep track of references to all aspects of the transport. Take notes on contradictory information to enable yourself or others to recapitulate decisions at a later stage. Identify all existing transport mechanisms. Keep the number of diffusion reactions as low as possible by resolving as many transport mechanisms and as detailed as possible. When using information from databases, confirm the correctness of the information based on the primary literature, e.g., confirm that it is extracellular transport and appears in your cell or organism.

4) **Add reactions**
Reconstruct the transport reactions based on your notes and add them to the model using rBioNet [46] in Matlab. Add missing exchange reactions using rBioNet [46].

**Stage 2: Set constraints and derive contextualized models**

5) **Define basic constraints, Trouble Shooting**
Use the function *setMediumConstraints* to impose the condition-specific sets of constraints to a model (*model*). When data values exceed the infinite bounds, increase the current bounds (*current_inf*) such that the infinite bounds no longer act as constraints (*set_inf*). Based on cell concentration (*cellConc*), the duration (*t*) of the cultivation, and the cell dry weight (*cellWeight*) fluxes are calculated from metabolite concentrations given in mM (*met_Conc_mM*) for a defined set of metabolites exchange reactions (*medium_composition*), which were part of the defined cell culture medium, and the model is constrained with these fluxes. Uptake of additional compounds, e.g., ions or vitamins can be restricted (*mediumCompounds*, *mediumCompounds_lb*). The optional input variable *customizedConstraints* allows individual definition of additional reaction constraints, e.g., growth rates.

>> [modelMedium, basisMedium] = setMediumConstraints(model, set_inf, current_inf, medium_composition, met_Conc_mM, cellConc, t, cellWeight, mediumCompounds, mediumCompounds_lb, [customizedConstraints], [customizedConstraints_ub], [customizedConstraints_lb], [close_exchanges])

6) **Transform the LOD from ng/ml to mM**

Use the molecular weight (*theo_mass*) of the metabolites and the respective exchange reaction for the metabolites (*ex_RXNS*) to transform the LOD from ng/ml (*lod_ngmL*) to mM (*lod_mM*).

\>> [LODmM] = calculateLODs(ex_RXNS, theo_MASS, LODngmL);

### 7) Define uptake and secretion profiles (no absolute quantification)
Define sets of metabolites that are consumed (*cond1_uptake* and *cond2_uptake*) and released (*cond1_secretion* and *cond2_secretion*) based on the individual data matrices (*input_A* and *input_B*) and the corresponding exchange reactions (*data_RXNS*) using the function *defineUptakeSecretionProfiles* for sample pairs. Manually define the threshold (*tol*) to accept a change as uptake or secretion. The function contains further options to manually tailor the uptake (*exclude_upt*, *add_upt*) and secretion profiles (*exclude_secr*, *add_secr*).

\>> [cond1_uptake, cond2_uptake, cond1_secretion, cond2_secretion, slope_Ratio, data_RXNS] = defineUptakeSecretionProfiles(input_A, input_B, data_RXNS, tol, essAA_excl, exclude_upt, exclude_secr, add_secr, add_upt);

### 8) Calculate semi-quantitative differences
Use the function *calculateQuantitativeDiffs* to add the LODs to the uptake and secretion profiles and to obtain the values for the relative adjustment of constraints (*cond1_upt_higher*, *cond2_upt_higher*, *cond2_secr_higher*, *cond1_secr_higher*) based on the *slope_Ratio* of the commonly consumed and released metabolites (Figure 5A). Additional inputs are the uptake and secretion profiles generated in the previous steps by the functions *defineUptakeSecretionProfiles* and *calculateLODs*. The input *LODmM* is of the same length as the vector specifying the exchange reactions *ex_RXNS*.

\>> [cond1_upt_higher, cond2_upt_higher, cond2_secr_higher, cond1_secr_higher, cond1_uptake_LODs, cond2_uptake_LODs, cond1_secretion_LODs, cond2_secretion_LODs] = calculateQuantitativeDiffs(data_RXNS, slope_Ratio, ex_RXNS, lod_mM, cond1_uptake, cond2_uptake, cond1_secretion, cond2_secretion);

### 9) Set qualitative constraints
Use the function *setQualitativeConstraints* to apply individual uptake (*cond_uptake*) and secretion (*cond_secretion*) profile to a model by using the detection limits to enforce uptakes and secretions (*cond1_uptake_LODs*, *cond2_uptake_LODs*, *cond1_secretion_LODs*, *cond2_secretion_LODs*). Execute this function for each model individually:

\>> [modelLOD] = setQualitativeConstraints(model, cond_uptake, cond_uptake_LODs, cond_secretion, cond_secretion_LODs, cellConc, t, cellWeight, ambiguous_metabolites, basisMedium);

Use the input variable *ambiguous_metabolites* to keep exchanges open and let the model freely consume or release the associated metabolites, e.g., if direction of exchange is undefined or differs between biological replicates. All lower bounds of exchange reactions, apart from those defined as ambiguous, mediumCompounds or from the uptake and secretion profiles, will be constrained to zero. The output model *modelLOD* has the same size of the input model.

**10) Apply semi-quantitative differences**

Use the function *setSemiQuantConstraints* to apply the semi-quantitative differences to two models *modelA* and *modelB*, which were previously defined by the function *calculateQuantitativeDiffs* (*cond1_upt_higher*, *cond2_upt_higher*, *cond2_secr_higher*, *cond1_secr_higher*). This step is performed simultaneously for the two models.

>> [modelA_QUANT,modelB_QUANT] = setSemiQuantConstraints(modelA, modelB, cond1_upt_higher, cond2_upt_higher, cond2_secr_higher, cond1_secr_higher);

**11) Obtain uptake and secretion profiles [integration of absolute concentration changes], Trouble Shooting**

Use the function *prepIntegrationQuant* to generate individual uptake and secretion profiles from a data matrix (*metData*) of fluxes, where the rows are the metabolite exchanges (*exchanges*) and the columns are the conditions (*samples*). Negative values are interpreted as uptake and positive values are interpreted as secretion. The individual exchange profile is saved to the user-defined location (*path*). The minimal and maximal flux values (*test_max*, *test_min*) are used to test whether the *model* can consume and secrete the *exchanges*. All fluxes below a user-defined tolerance value (*tol*) are set to zero. The output consists of the flux values to be applied to the upper and lower bounds, which are based user-defined variation (of, e.g., 20%, *variation*) for uptake and secretion, and written to a file named according to the same name to the specified location.

>> prepIntegrationQuant(model, metData, exchanges, samples, test_max, test_min, path, tol, variation);

**12) Check the exchange profiles**

Check if additional metabolites have been removed from the uptake and secretion profiles. Use the function *checkExchangeProfiles* to generate statistics on the number and identity (*mapped_exchanges*, *mapped_uptake*, and *mapped_secretion*) of uptake and secretions added per sample (*minMax*). Specify the number of metabolites in the data set (*nmets*). The function loads the exchange profiles generated by using the definition of the *path* and the sample names (*samples*).

>> [mapped_exchanges, minMax, mapped_uptake, mapped_secretion] = checkExchangeProfiles(samples, path, nmets);

**13) Integrate quantitative constraints, Trouble Shooting**

Use the function *setQuantConstraints* to generate a contextualized model (*ResultsAllCellLines.sample.modelPruned*) for each sample (*samples*). A minimal flux value (*minGrowth*) can be specified for a defined objective function (*obj*). A set of metabolite exchanges (*medium*) can be specified, which will be part of the retained minimal set of exchange reactions in the output model. Allow additional secretion of metabolites (*addExtraExch*, *addExtraExch_value*). The exchange profile is loaded from the specified location (*path*). The default value of epsilon is $1e^{-4}$, see [26] for a detailed description.

>> [ResultsAllCellLines,OverViewResults] = setQuantConstraints(model, samples, tol, minGrowth, obj, no_secretion, no_uptake, medium, addExtraExch, addExtraExch_value, path, [epsilon]);

The output structure *ResultsAllCellLines* contains the pruned and unpruned but constrained contextualized model for each sample (*samples*). *OverViewResults* is a table which lists statistics of the generated models.

**14) Generate the contextualized subnetworks**

A) Use *generateCompactExchModel* to generate a submodel. This pruned model is based on the prediction of a minimal set of metabolite exchanges for a given, constrained model, e.g., the model resulting from running *setQualitativeConstraints* or *setSemiQuantConstraints*, while permitting a feasible steady state flux distribution. The prediction of the additional metabolite exchanges is based on the minimization if the cardinality of the exchange reactions [32]. Define the minimal flux value (*minGrowth*) that should be achieved by a defined objective (*biomassRxn*) in the output model (*modelMin*, *modelPruned*). The use of fast flux variability analysis [60] (*fastFVA*) is optional.

>> [modelMin, modelPruned, Ex_Rxns] = generateCompactExchModel(model, minGrowth, biomassRxn, prune, fastFVA);

B) Alternatively, use *extractConditionSpecificModel* to extract the submodel using the approach as in [7]. Specify the cutoff (*theshold*) for calling a flux value zero (e.g., $1e^{-8}$).

>> [modelPruned] = extractConditionSpecificModel(model, theshold);

**15) Integrate gene expression data**
Use the function *integrateGeneExpressionData* to integrate transcriptomic data (set of genes found to be unexpressed (*dataGenes*)) with the *model*.

>> [modelGE] = integrateGeneExpressionData(model, dataGenes);

**Stage 3: Validation and prediction**

**16) Validation of the model, Trouble Shooting**
Perform a quality control of the models by checking constraints against biological insight. The steps 16E-F can only be performed if the data exists.

A) **Create an expectation**
If not done already, review the literature to learn about the biological background of your cells, organisms, and what to expect from the analysis: Identify metabolic traits and inabilities of the target cell and what is known about the respective set of environmental conditions and its response to perturbations. Validate the generated models based on these biological insights, e.g., by asking questions such as those given in Table 3.

B) **Check constraints, Trouble Shooting**
Check lower and upper reaction bounds in the generated models. Check that all constraints have been applied as intended, e.g., typos in the exchange reaction names might have prevented that constraints were applied. Are all exchange reaction constrained to zero that should have been closed? Do constraints exceed the infinite bounds? Are any constraints below the threshold and thus equal to zero? Are there any constraints that do not concur with the expectation, e.g., uptake of oxygen?

C) **Analyze added exchanges, Trouble Shooting**
Summarize the set of added exchanges using the functions *statisticsAddedExchanges* and *mkTableOfAddedExchanges*:

>> [Ex_added_all_unique] = statisticsAddedExchanges(ResultsAllCellLines, samples);

>> [Added_all] = mkTableOfAddedExchanges(ResultsAllCellLines, samples, Ex_added_all_unique);

Check the additional exchanges that were added to the model. Are the metabolites predicted to be consumed part of the medium composition? Was serum added to the culture medium? For all metabolites, which the model needs to secrete, could these metabolites have been detected? Were they targeted by the metabolomic analysis? Could they have been below the limit of detection? Does untargeted metabolomic data support the presence of these metabolites in the extracellular medium?

D) **Check metabolic functions, Trouble Shooting**
After confirming that the constraints are accurately set, check the functionality of the model against the expectation build earlier. Does the model include metabolic pathways known to be operated by the cell and in the considered condition? Which pathways are in the model that are not expected to be present in the modeled cell type? Keep in mind that the model was optimized to explain the uptake and secretion profile and that the internal reaction redundancy remains, if no additional omics data is included. Check, using flux variability analysis, if the unexpected pathways have to carry flux. What imposed exchanges are responsible for the flux through such a pathway?

E) **Validate the contextualized model based on experimental data**
Use, e.g., intracellular metabolomic data. Check to which extend metabolites reported in intracellular metabolomic data are captured by the contextualized model (i.e., are the metabolites part of model.mets). Do the models reflect sample-specific differences?

F) **Test for condition-specific growth rates**
If growth rates were not applied as constraints during the model building, use the function *setConstraintsOnBiomassReaction* to apply constraints to the biomass reaction based on doubling times (*dT*). The upper and lower bounds of the biomass objective function (*of*) are adjusted with a *tolerance* around the calculated growth rate (e.g., 10 %).

>> [modelBM] = setConstraintsOnBiomassReaction(model, of, dT, tolerance);

Alternatively, use the function *changeRxnBounds* to apply growth rates to the model. Make sure the objective is set in the model (use the function *changeObjective* to change the objective in the model).

If no doubling times or growth rates exist for the data set, literature derived rates can provide an approximation for the expected growth rates, as growth rates can vary depending on the culture conditions (e.g., serum or no serum). Use flux balance analysis (*optimizeCbModel*) to see if the model can produce biomass at experimental rates.

17) **Predict metabolic phenotypes, Trouble Shooting**

Choose the most suitable analysis (e.g., 17A-E) based on the individual biological question to gain insight into the use of the most relevant pathways, reactions, or metabolites by individual or groups of models.

A) **Summarize and illustrate the set of essential genes**

Use the function *analyzeSingleGeneDeletion* to define and to analyze the set of essential genes for single or large set of models (*ResultsAllCellLines*, *samples*). Define through the input variable *heat* if a heat map should be generated, illustrating the differences across the model set (Figure 6). Save the output at the specified location (*path*).

>> [genes,ResultsAllCellLines,OverViewResults] = analyzeSingleGeneDeletion (ResultsAllCellLines, path, samples, cutoff, OverViewResults);

The output variable *genes* contains a table that lists per unique gene the number of models with 'no effect', 'partial effect', essential genes or 'KO', and in how many models the gene is not present. Additionally, the number of essential genes is added to *OverViewResults* and the output of the single gene deletion is added to *ResultsAllCellLines*.

B) **Identify the essential reactions driving gene essentiality, Trouble Shooting**

Use the function *checkEffectRxnKO* to identify, which reactions are essential. Provide the set or subset of models (*samples_to_test*) from *samples* that should be tested. The functions tests all individual reactions associated with a set of specified genes (*genes_to_test*). The variable *fill* defines what to add into the output matrix if a reaction is not in the model, e.g., 100, num('NAN')).

>> [FBA_Rxns_KO, ListResults] = checkEffectRxnKO(samples_to_test, fill, Genes_to_test, samples, ResultsAllCellLines);

The output variable *FBA_Rxns_KO* contains the FBA results for constraining one reaction at a time to zero. The second output *ListResults* lists the reactions associated with the input gene list in the same order as *FBA_Rxns_KO*.

C) **Generate intersect and union model**

Use the function *makeSummaryModels* to find the shared and superset of reactions, metabolites and genes of a set of models by creating a union (*unionModel*) and an intersect model (*intersectModel*). The models are generated based on the generic model (*model*) from which the subnetworks have been generated.

```
>> [unionModel, intersectModel, diffRxns, diffExRxns] =
makeSummaryModels(ResultsAllCellLines, samples, model, mk_union,
mk_intersect, mk_reactionDiff);
```

Both output models (*unionModel*, *intersectModel*) and differential reaction sets (*diffRxns*, *diffExRxns*) are generated by default.

D) **Predict flux splits**

Use the function *predictFluxSplits* to predict either production or consumption (*dir*) of metabolites (e.g., *met2test* = {'atp'}) for the models in *ResultsAllCellLines* defined by *samples*. Define the objective function (*obj*) that should be used to generate the flux vector, from which the reaction contributions to production or consumption of the considered metabolite are calculated. Define a full sized model (*model*), as the set of models can use distinct subsets of the reactions of the full model to produce or consume the metabolite of interest. A default value (eucNorm = $1e^{-5}$) is set for the parsimonious flux balance analysis. Set a quadratic programming solver (*changeCobraSolver*(solver,'QP')) before running *predictFluxSplits*.

```
>> [BMall, ResultsAllCellLines, metRsall, maximum_contributing_rxn,
maximum_contributing_flux, ATPyield] = predictFluxSplits(model, obj, met2test,
samples, ResultsAllCellLines, dir, transportRxns, ATPprod, carbon_source,
eucNorm);
```

Exclude reactions, which do not produce or consume the metabolite, e.g., reactions that transport the metabolite between compartments. Define all reactions that produce/consume the metabolite in a first run (*transportRxns* = []), and subsequently, exclude unwanted contributors by adding them to *transportRxns*.

In case that the production of ATP is predicted, additional statistics on the use of glycolysis, the citric acid cycle, and the oxidative phosphorylation can be automatically generated (*ATPprod*=1). With this option, the *ATPyield* is calculated, which relates the total ATP production to a user-defined carbon source (default: *carbon_source* = 'EX_glc(e)';).

The output lists the reaction with highest flux value for producing or consuming a defined metabolite across analyzed samples (*maximum_contributing_rxn*, *maximum_contributing_flux*) to simplify the analysis over large groups of samples (and metabolites of interest). Detailed results of the analysis are added to *ResultsAllCellLines*.

E) **Sampling the steady-state solution space**

Use the function *performSampling* to sample the solution space of a model:

```
>> performSampling(model, warmupn, fileName, nFiles, pointsPerFile,
stepsPerPoint, fileBaseNo, maxTime, path);
```

The function is based on COBRA toolbox, refer to the corresponding protocol for details [29]. Subsequently, use the function *summarizeSamplingResults:*

```
>> [stats, statsR] = summarizeSamplingResults(modelA, modelB, path, nFiles,
pointsPerFile, starting_Model, dataGenes, show_rxns, fonts);
```

Inspect the histograms automatically saved to the user-defined location (*path*, Figure 8, supplemental material). If the distributions are not unimodal (only bounded reactions will reach this, but not loop reactions (see Box 1), generate more sample points by using points from the last file as starting point (see tutorial_I). Analyze the results: Are the distributions shifted apart for any reactions or pathways that were expected or unexpected for a certain condition or group of samples? Make sure that your interpretations do not rely on unbounded reactions (Figure 8).

F) **Predict behavior to environmental changes**

Use the function *performPPP* to predict the simultaneous variation of two parameters (*mets*) at a time and for one or more submodels (*samples* in *ResultsAllCellLines*). Define the range of flux rates (uptake or secretion defined in *direct*) through the variables *step_size* (flux units) and number of steps *step_num*. This can be individually defined for each exchange reaction. The output is added to *ResultsAllCellLines*.

```
>> [ResultsAllCellLines] = performPPP(ResultsAllCellLines, mets, step_size,
samples, step_num, direct);
```

Illustrate your results with heat-maps using the function *illustrate_ppp*.

```
>> illustrate_ppp(ResultsAllCellLines, mets, path, samples, label, fonts, tol);
```

The heat maps are saved to the defined *path*. The function uses flux variability analysis [29, 60], and the zero cutoff (*tol*, e.g., $1e^{-8}$) needs to be specified. The function allows some additional inputs, i.e., label and font size (*label*, *fonts*), to customize the illustrations. Analyze the output, e.g., by manually grouping the models based on the similarity of the planes in the heat maps (e.g., size and shape).

18) **Validate the predicted, metabolic phenotypes**

A) **Compare to experimental data**
Compare the results of the previous steps to data sets, e.g., cell culture and perturbation experiments, or compare the predictions to the results of the analysis of additional omics data, e.g., mRNA. Does regulation in metabolic genes support predicted metabolic differences between cells or conditions?

B) **Compare to literature**
Check if the results of the predictions generated in the previous step agree with what is reported in the literature. For example, has the cell line been previously reported to heavily rely on an alternative pathway, e.g., reverse flux through the citric acid cycle? Are the essential genes already known to be essential in this cell type, or under the experimental conditions, such as hypoxia?

5. **Trouble Shooting**

**Step 1: Metabolites cannot be associated with model metabolites.** Exclude metabolites, which are not part of the model. Leaving out constraints has as consequence that physiologically irrelevant flux distributions (network states) may remain part of the steady-state solution space. However, no wrong flux distributions are introduced! Another solution would be to add demand reactions (see Box 1) [43] that 'artificially' remove metabolites from the system, which were experimentally consumed but that could not be consumed by the model. A sink reaction (see Box 1) [43] could be included to represent an intracellular source for a metabolite, which cannot be produced by the model. However, such reactions have only an effect on the model if, e.g., the transport of the metabolite is dependent on energy or other substrates and might otherwise also be ignored. The consecutive addition of demand and sink reactions is not recommended.

**Step 2: Exchange reactions are constrained although the formula indicates that they are reversible.** Change reaction bounds on the transport reaction using the function *changeRxnBounds*. Reactions might be closed to prevent thermodynamically infeasible cycles. Opening these reactions might lead to a leaking model (i.e., the model is able to produce metabolites, such as ATP without substrates). Thus, the model should be tested for leakage before proceeding to further steps (see [43]).

**Step 3: Transporter is unknown.** Add a diffusion reaction.

**Step 4: Existing reactions need to be modified. New substrates have been identified for an existing transporter.** Add, replace, or modify reactions if literature provides novel, convincing clues on the metabolite transport mechanisms. If diffusion is unlikely for a metabolite, consider removing the diffusion reaction that allow the transport 'for free'. Add a reaction if the metabolite constitutes an additional substrate to an existing transporter. Make sure that the new substrate is transported in the same way as the original one before duplicating the reaction. Does the reaction mechanism correspond to that in the model? Check if the reaction stoichiometry, directionality, co-substrates, genes, and Gene-protein-reaction association (GPR) of the existing reaction agrees with your notes. Add the reaction with a new name corresponding to the transported metabolite. If the transport mechanism is different from the reaction mechanism already present in the model (particularly for diffusion reactions), review whether the original reaction needs to be updated based on the novel literature clues. The mechanism might have been better resolved since the original reconstruction (see references associated with the reaction, e.g., VMH database for the human genome scale reconstruction, https://vmh.uni.lu/). Correct the original reaction in the model based on the result of the literature review. Remove diffusion reactions if no longer adequate.

**Step 5: Identify minimal set of exchanges for a model and objective.** Use flux variability analysis (FVA) to identify mandatory metabolite uptakes (negative non-zero minimal and maximal flux).

**Step 7: Detection limits are unavailable.** An arbitrary low value may be used to set the upper bounds on uptake fluxes and lower bounds on secretion fluxes that is close to the cutoff value (e.g., -0.00001/0.00001).

**Step 13: Generate metabolic fluxes.** Use the function *conc2rate* to convert concentration changes (*metConc*) measured over the time course of the experiment (*t*), for a certain number of cells (*cellConc*), and dry weight (*cellWeight*) into flux (*flux*).

>> [flux] = conc2Rate(metConc, cellConc, t, cellWeight);

**Step 16B: Constraints are too small and fall below the error range and/or exceed the infinite bounds (COBRA toolbox default is $1e^{-8}$).**
Multiply or divide all reaction bounds (except infinite constraints) with the same factor. Adjust the unit the fluxes accordingly using the same factor. Scale the unit the fluxes are reported in with the same factor. Repeat the previous steps to apply the reconciled constraints. If scaling is not possible because the range of fluxes is too wide (e.g., -1000 to 1000), the infinite bounds can be increased (*setMediumConstraints*).

**Step 16C: Additional metabolite exchanges are unlikely from a biological point of view.**
Close exchange reaction in the starting model and repeat the model building by *setQuantConstraints* (or *generateCompactExchModel*).

**Step 16D: The model contains unexpected pathways.** Keep in mind that the integration of metabolomics data only indirectly causes the removal of metabolic pathway. Pathways are only removed if they become blocked as a consequence of exchanges being closed. An additional reduction of the internal metabolic network can be achieved by the integration of additional omics data sets, if the imposed flux rates are not deterministic enough. **The contextualized models miss central features of the cells or certain metabolic functions cannot be performed.** Note that the models constructed herein are condition-specific. Are the reported pathways really active in the considered experimental condition? Identify whether there is an exchange connected to the pathway, which is not defined by the consumed metabolites. Could these be missing from set of targeted metabolites? You can add the metabolite to the medium composition (e.g.,*ambiguous_metabolites*) to prevent that it is removed during model building. Iterate until models are conform to the expectation.

**Step 17: Generate the variables *ResultsAllCellLines* and *OverViewResults* to collect the information of the downstream analysis for models not generated herein.** For models of name *model*, *model2*, ... loaded in the workspace, execute:

```
>> samples = {'model', 'model2', …};
>>   for i=1:length(samples)
>>     model_name= samples{i}
>>     eval(['ResultsAllCellLines.' model_name '.modelMin =' model_name]);
>>     eval(['ResultsAllCellLines.' model_name '.modelPruned =' model_name]);
>>   end
```

The difference between the models is that the *modelMin* has all reactions, whereas the pruned model is the reduced model. If only one exist, both can be equal. In the workflow the *modelMin* is used for the single gene deletion, instead of the *modelPruned*.

To generate the variable *OverViewResults*, execute:
```
>> samples = {'model', 'model2'};
>> cntO=1;
>> OverViewResults{1,cntO} ='samples';
>> for i=1:length(samples)
>>     OverViewResults{i+1,cntO} = samples{i};
>> end
```

Some of the analysis require the specification of a generic model, which is a superset of the models specified in *samples*.

**6. Timing**

Step 1, Map metabolite identifiers to Recon IDs, Timing: ~hours
Step 2, Find which metabolites cannot be transported in the model, Timing: ~minutes
Step 3, Identify missing metabolite transporter, Timing: depends on number ~ hours - days
Step 4, Add reactions, Timing: ~hours
Step 5, Define medium constraints, Computational Time: ~seconds
Step 6, Calculate detection limits, Computational Time: ~seconds
Step 7, Define uptake and secretion profiles, Computational Time: ~seconds
Step 8, Calculate semi-quantitative differences, Computational Time: ~seconds
Step 9, Set qualitative constraints, Computational Time: ~seconds
Step 10, Apply semi-quantitative differences, Computational Time: ~seconds
Step 11, Obtain uptake and secretion profiles (flux rates), Computational Time: ~seconds
Step 12, Check the exchange profiles, Computational Time: ~seconds
Step 13, Integrate quantitative constraints and generate minimal exchange model, Computational Time: ~minutes per model
Step 14, Generate the subnetwork by pruning model, Computational Time: ~minutes
Step 15, Integrate gene expression data, Computational Time: ~seconds

Step 16, Validate the model, Timing: ~ days to weeks
  A) Literatue review to create an expectation, Timing: ~ days to weeks
  B) Check constraints, Timing: ~ minutes to hours
  C) Analyze added exchanges, Timing: ~ minutes to hours
  D) Check metabolic functions, Timing: ~ minutes to hours
  E) Comparison to experimental data, Timing: ~ days to weeks
  F) Test growth rates, Computational Time: ~ seconds

Step 17, predict metabolic phenotypes, Timing: ~ hours to weeks
  A) Define essential genes, Computational Time: ~ minutes
  B) Identify essential reactions behind the gene essentiality, Computational Time: ~seconds
  C) Generate intersect and union model, Computational Time: ~seconds to minutes
  D) Predict flux splits, Computational Time: ~seconds per model
  E) Sampling the steady-state solution space, Computational Time: ~hours to days per model
  F) Predict behavior to environmental changes, Computational Time: ~minutes to hours per model

Step 18, Validate predicted, metabolic phenotypes, Timing: ~ days to weeks (or months)

**Acknowledgments**
We thank Alberto Noronha, Dr. Nathalie Poupin, and Dr. Averina Nicolae for testing and commenting on the tutorials and the manuscript.

**Funding**
This work is supported by the Luxembourg National Research Fund (FNR) through the National Centre of Excellence in Research (NCER) on Parkinson's disease, by the FNR ATTRACT program grant (FNR/A12/01; IT and MA). RF is funded by the U.S. Department of Energy, Offices of Advanced Scientific Computing Research and the Biological and Environmental Research as part of the Scientific Discovery Through Advanced Computing program, grant #DE-SC0010429. This project has also received funding from the European Union's Horizon 2020 research and innovation programme under grant agreement No 668738.

**Author Contributions**
Contributed code to MetaboTools: IT, MA, RF. Designed the tutorials: MA. Wrote the paper: MA, RF, and IT.

**Competing financial interests**
The authors have no competing financial interests.
**References**
1. Cuperlovic-Culf, M., et al., *Cell culture metabolomics: applications and future directions.* Drug Discov Today, 2010. **15**(15-16): p. 610-21.
2. Bordbar, A., et al., *Model-driven multi-omic data analysis elucidates metabolic immunomodulators of macrophage activation.* Mol Syst Biol, 2012. **8**: p. 558.
3. Bordbar, A., et al., *Insight into human alveolar macrophage and M. tuberculosis interactions via metabolic reconstructions.* Mol Syst Biol, 2010. **6**: p. 422.
4. Mo, M.L., B.O. Palsson, and M.J. Herrgard, *Connecting extracellular metabolomic measurements to intracellular flux states in yeast.* BMC Syst Biol, 2009. **3**: p. 37.
5. Allen, J., et al., *Discrimination of modes of action of antifungal substances by use of metabolic footprinting.* Appl Environ Microbiol, 2004. **70**(10): p. 6157-65.
6. Krug, S., et al., *The dynamic range of the human metabolome revealed by challenges.* FASEB J, 2012. **26**(6): p. 2607-19.
7. Aurich, M., et al., *Prediction of intracellular metabolic states from extracellular metabolomic data.* Metabolomics, 2015. **11**(3): p. 603-619.
8. Petersen, A.K., et al., *Epigenetics meets metabolomics: an epigenome-wide association study with blood serum metabolic traits.* Hum Mol Genet, 2013.
9. Saito, K. and F. Matsuda, *Metabolomics for functional genomics, systems biology, and biotechnology.* Annu Rev Plant Biol, 2010. **61**: p. 463-89.
10. Kell, D.B., *Metabolomics and systems biology: making sense of the soup.* Curr Opin Microbiol, 2004. **7**(3): p. 296-307.
11. Orth, J.D., I. Thiele, and B.O. Palsson, *What is flux balance analysis?* Nat Biotechnol, 2010. **28**(3): p. 245-8.
12. Joyce, A.R. and B.O. Palsson, *The model organism as a system: integrating 'omics' data sets.* Nat Rev Mol Cell Biol, 2006. **7**(3): p. 198-210.
13. Bordbar, A. and B.O. Palsson, *Using the reconstructed genome-scale human metabolic network to study physiology and pathology.* J Intern Med, 2012. **271**(2): p. 131-41.
14. Frezza, C., et al., *Haem oxygenase is synthetically lethal with the tumour suppressor fumarate hydratase.* Nature, 2011. **477**(7363): p. 225-8.
24

**Figure Legends**

**Figure 1: This protocol and the MetaboTools provide comprehensive support for integration of extracellular metabolomic data and for the analysis and phenotypic stratification of metabolic models.** The workflow captures the integration of processed and annotated metabolomic data sets, the generation of contextualized submodels, and the computational model analysis to distinguish metabolic phenotypes. The individual divided into small sections and form a step-by-step procedure discussed in detail in the protocol. The tutorials (supplementary tutorial I & II) demonstrate two use-cases of the workflow and MetaboTools.

**Figure 2: Stage 1 - Preparations for the integration of the metabolomic data. A.** First, the detected metabolites need to be matched to the metabolite ID used in the model. This step is simplified if the metabolomic data are annotated with standard metabolite IDs (e.g., from KEGG [39] or HMDB [41]). **B.** Check if transport and exchange reactions exist in the model for each of the matched metabolites using the model metabolite ID. In case of the human genome-scale reconstruction (Recon 2 [35]), you can check for reactions using the virtual metabolic human database (VMH, https://vmh.uni.lu/) or directly in your model with the COBRA toolbox function *findRxnsFromMets*. The illustrated VMH query for glutamine shows two of the glutamine transport reactions (highlighted in blue) and the exchange reaction (highlighted in green) of glutamine. Exchange reactions are by definition written as in the given example. **C.** Transport reactions need to be identified as comprehensively as possible from the literature. Based on the information gathered about the transport mechanisms, by which a metabolite is transported into and out of the cell, the transport reactions can be formulated. The transport and exchange reactions can be added to the model using rBioNet [46].

**Figure 3: Shaping the model to predict condition-specific metabolic states (contextualized model). A**. In order for the model to follow the measured metabolite uptake and secretion profile, both exchange reactions and transport reactions need to be present to force the metabolite into (uptake) and out of the model (secretion). If those do not exist in the model, the transporters along with the mechanisms need to be identified in order to allow the model to follow the uptake and secretion profile of the experimental data. **B**. The uptake of metabolites is constrained based on the concentrations of metabolites available to the conditions that should be investigated. For example, the composition of defined experimental medium can be used. Also, the maximum uptake per cell and time unit can be calculated from the concentrations of nutrients, the concentration of cells, the cell weight, and the duration of the experiment. Otherwise, those constraints would be infinite and relative differences in the metabolite uptake would be less bound to the actual environmental conditions one wants to investigate. **C.** Metabolomics data are mapped as constraints to the exchange reactions. Glutamine, glyceraldehyde, and citrulline have to be secreted by the model. **D.** Transcriptomic or proteomic data are mapped to the internal network reactions (qualitatively), restricting which reactions a cell can use to transport and metabolize the metabolize, e.g., the enforced secretion of glutamine.

**Figure 4: Application of constraints on the bounds of exchange reactions.** Infinite bounds (here, -1000 to 1000) express the unlimited supply and removal of metabolites to and from the system. A negative flux through an exchange reactions corresponds to an uptake (red) and a positive flux corresponds to the secretion of a metabolite (blue). Maximal uptake can be constraint by restricting the lower bound (lb = -600). The directionality of the exchange can be defined by setting a negative value as the upper bound (ub) for an uptake or by setting a positive value as the lower bound for secretion. These constraints could be based on the limit of detection (LOD) or quantification for the metabolite. Relative differences in the uptake and secretion of a metabolite can be expressed as the relative difference from maximal possible uptake rate (purple as compared to red) and as the relative difference from the minimal secretion rate. Absolute concentration changes over time can be converted to flux values and added to the constraints, considering a user-defined error to the flux to define lower bound and upper bound.

**Figure 5: Applying semi-quantitative constraints.** Relative differences in the uptake and secretion of metabolites from and into the extracellular medium can be imposed on the models, emphasizing the differences in metabolite uptake and secretion between samples. **A.** A slope ratio is calculated based on the data as the relative difference in uptake of a metabolite between two samples and compared to the control (medium). Flux values are in the unit of mmol/$g_{dry\ weigth}$/hr. The result of the preparation is a set of new lower bounds, which can be applied to the model. Note that the adjusted bound is the grey number, whereas the orange number constitutes the flux value, which was defined based on the medium composition (negative bounds for maximal possible uptake) or based on the minimal detection limits (positive numbers for minimal possible secretion). **B.** Illustration of the trend in quantitative differences glucose between cell lines. A difference in the measured intensity can be observed between time point 0 and 48 hr. The signal also decreases in the control (medium). **C.** Relative differences of metabolite uptake and secretion are translated into relative differences of the constraints on exchange reactions of the cell line models, forcing the models to consume or release metabolites in the same relation as observed experimentally.

**Figure 6: Heat-map depicting differences in single gene deletion across 120 NCI-60 cell line models.** The growth ratio is defined as the maximal objective value of the model divided by the maximal objective value of the model when the gene was deleted when optimizing for biomass production. Genes that were absent in individual models have a growth ratio value of -1. Genes that were absent in all models and genes whose deletion did not affect any model are not illustrated. The coloring in the heat-map corresponds to the maximal objective values obtained from flux balance analysis at each step of the analysis.

**Figure 7: Models can be stratified based into distinct metabolic phenotypes on the pathway that they use prevailingly.**

**Figure 8: Interpreting the results of the sampling analysis. A.** During the sampling analysis, a large set of sampling points is collected (red dots, each comprising a flux distribution). The fluxes collected from the sampling points can be illustrated as probability distribution for individual reactions. **B.** Comparing the histograms of the probability flux distributions through the glutamine exchange reaction in two models (blue and green) reveals a shift of the feasible steady-state solution space for this reaction. **C.** The irregular histogram is typical for an unbounded reaction (distribution ranges from -1000 to 1000). **D.** A premature flux distribution of the glutamine exchange reaction. Even though the final shape of the distributions can be guessed, additional sampling points need to be generated to get a unimodal distribution of the fluxes. Generation of additional points could shift the distribution. U= flux units.

Tables
Table 1: Additional data used for the intra-model analysis of metabolomic data.

| Data type | Unit | Use case I: Semi-quantitative data | Use case II: Quantitative data |
|---|---|---|---|
| **Cell medium composition (e.g., RPMI medium)** | mM | yes | no |
| **Concentration of Ions, and other compounds (e.g., $O_2$, $CO_2$)** | mM | $O_2$ defined based on literature. | $O_2$ defined based on literature for one of the cell lines. |
| **Cell-weight** | preferable $g_{dry\ weight}$ | experimental measurement | defined based on literature for one of the cell lines |
| **Cell count** | cells per ml | experimental measurement | not required |
| **Experimental duration (in hours)** | | experimental duration | not required |
| **Detection limits for all detected metabolites (mass spectrometer)** | instrumental limit of detection in ng/mL or LODs in mM | experimental measurement | not required |
| **Doubling time** | hours or growth rate | experimental measurement | defined in the experiment if possible |
| **Transcriptomic or proteomic data** | list of absent genes | experimental measurement from the same experiment | not required |

Table 2: Test case illustrating possible loophole in the reaction constraints.

| Test case: | Missing constraints |
|---|---|
| **Expectation** | A human model requires oxygen, no constraints are applied to oxygen exchange. |

| **Reality** | Contextualized models secrete oxygen. |
|---|---|
| **Consequence** | Additional constraints need to be applied to prevent oxygen secretion. |

**Table 3: Use expectations to compare the generated models against.** The table lists examples of questions and data that can be used to validate contextualized models along with existing examples and MetaboTools functions that can be used for this analysis.

| Question to address for model validation | Example used in Leukemia cell lines [7] or NCI-60 cell lines [32] | Check using MetaboTools*, models, or alternative resource |
|---|---|---|
| What metabolic pathways does/ does my model include that would (not) be expected for the target cell? | Transcriptomic data integration caused absence of complex I of the electron transport chain in the models, which complied with literature [7]. | model.subsystems, model.rxns<br><br>metabolic functions [35] |
| To what extend does the model capture metabolites detected in the intracellular metabolome? | | model.mets |
| Are the models able to achieve experimental growth rates given the applied constraints? | -Analysis was conducted using biomass constraints [7].<br><br>-The vast majority of the models was able to grow at experimental growth rates [32]. | *setConstraintsOnBiomassReaction* **or** *changeRxnBounds* (add constraint)<br><br>*optimizeCbModel* (perform FBA)<br><br>*(changeObejctive* – set objective function in model.c*)* |
| Which exchange reactions have been added? Are the cells known to use/ secrete these substrates? | Cancer cells are known to use fatty acids to support their growth [32]. | *statisticsAddedExchanges** |

**Table 4: Questions to guide the analysis of predicted sets of essential genes.**

| Purpose | Question |
|---|---|
| Stratify models into groups | How much does the number of essential genes vary across models? Does this variation coincide with model size or growth rates, or with phenotypes predicted by another analysis? |

| | Which genes are essential only in a small subset of the submodels? Which essential genes affect all models in the set of model? |
|---|---|
| | Are the essential genes directly connected to an enforced metabolite exchanges? |
| Validation, generate testable hypothesis | Are the predicted essential genes known or already used for therapeutical purposes? |
| Generate testable hypothesis | Do you find interesting essential genes, pathways affected in single models or model groups? Could predicted essential genes constitute interesting targets (e.g., drug targets)? |

**Boxes**

**Box 1**
**Glossary**
**Antimetabolite** – structurally similar metabolite that inhibits the use of the original metabolite by an enzyme and thus, can diminish proliferation.
**Cell-type or tissue-specific model** – model that contains all reactions and pathways a cell type or tissue can use in under any set of environmental and/or genetic condition. A condition-specific model contains a subset of reactions of the cell-type or tissue-specific model.
**Closed exchange reaction** – reaction, whose lower and upper bounds are set to zero and which, due to these constraints, cannot carry any flux (flux value is zero).
**Constraints** – limits set on reactions in the metabolic model and defined through the upper bound ('ub') and lower bound ('lb') variable vector in the model. Possible computed flux values for a reaction can only lie between or on these bounds.
**Contextualized model** – model that has been tailored towards a particular experimental condition.
**Data integration** – applying constraints based on experimental (e.g., high-throughput) measurements.
**Data mapping** – matching names of the metabolites in the reconstruction and the data.
**Demand reaction** – unbalanced network reactions, which allow accumulation of metabolites in the model and as such circumvent the steady-state assumption underlying constraint-based modeling.
**Exchange profile** – the profile of a sample (e.g., cell line, hypoxic condition, etc.), detailing, which metabolites are consumed and which metabolites are released. It will be used to established constraints during the data integration process.
**Feasible model** – a model that can satisfy the user-defined objective function while being consistent with the applied constraints.
**Gap-filling approaches** – predict possible ways to fill network gaps using pathway and reaction information from other organisms.
**Generic model** – a model that captures all metabolic functions and pathways known to be active in at least one cell and/or under at least one condition in organism.
**Gene-protein-reaction association (GPRs)** - a reaction is associated with one or more genes encoding for one or more enzymes, which catalyze the reaction. The gene-protein-reaction associations are formulated as Boolean rules, whereby isozymes are associated with an OR and subunits of the functional protein complex are associated with an AND in the GPRs.
**Infinite constraint** – an arbitrary high value that does not limit reaction flux.

**Loop reactions** – cyclic internal reactions that do not depend on the inputs and outputs of the model. Loop reactions are thermodynamically infeasible and common artifacts in metabolic models due to missing constraints (e.g., temporal separation of pathways, regulation).
**Open model** – all exchange reactions in the model are unconstrained.
**Sample** – one experimental condition.
**Sampling point** – each sampling point contains an entire flux vector, and thus contains a flux value for each reactions in the metabolic model.
**Sink reaction** – adds compounds that cannot be produced by the metabolic model, because they originate from pathways outside metabolism or because the production pathways are unknown.
**Steady-state assumption** – the modeled system is assumed to be in steady state, i.e., no metabolites are accumulated over time ($dx/dt = 0$).
**Unbounded reactions** – reactions, whose fluxes are not limited through the network topology and applied constraints, but which expand over the entire span between the constraints (i.e., infinite bounds).
**Unconstrained** – the upper and lower bounds are set to infinite.

**Supplementary data**
1. **Supplementary material (text)**
    1.1 The model structure in matlab
    1.2 Infinite constraints
    1.3 Why use minExCard?
    1.4 Conversion of the theoretical mass
    1.5 Calculate weight and obtain dry weight
    1.6 Generate metabolic fluxes
    1.7 Additional on scaling of infinite bounds and defined constraints
    1.8 Additional remarks on the integration of Gene Expression data
    1.9 Additional remarks on "Sampling the solution space"

    **Supplementary Figure 1**: Structure in the human metabolic model.

    **Supplementary Table 1: Appeal from a computational biologist to the metabolomics community.** The table lists points that simplify the integration of metabolomics data sets into the model context.

2. **Supplementary Tutorial I**: Workflow for the integration of semi-quantitative extracellular metabolomic data into the network context.
3. **Supplementary Tutorial II**: Workflow for the integration of quantitative extracellular metabolomic data into the network context.

**Figures**

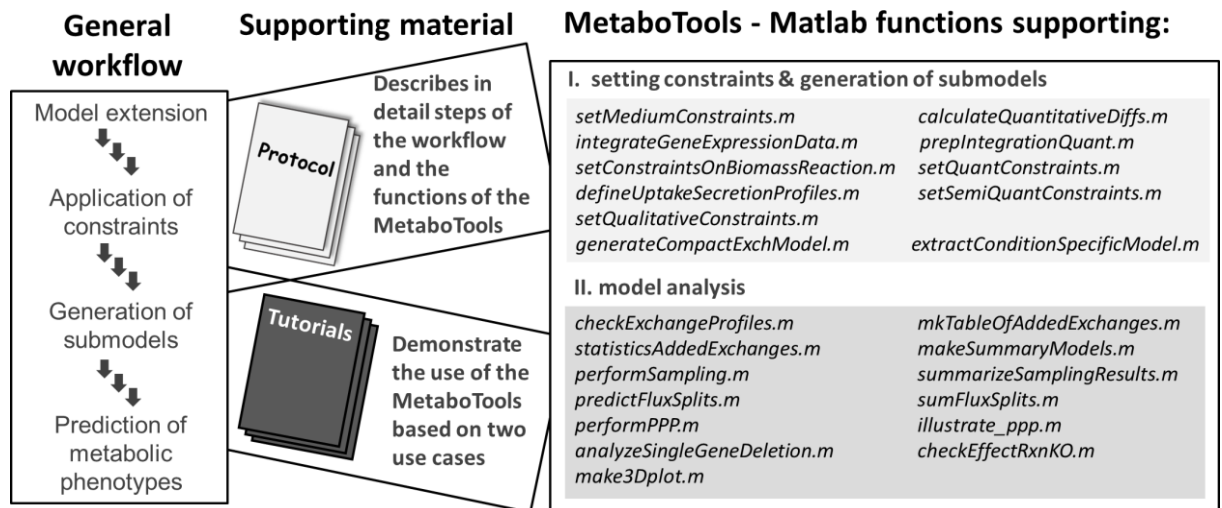

Figure 1

## A) Match metabolite identifiers between data and model

## B) Check model for existing transport and exchange reactions

| Model metabolite abbreviation | Transport reaction present ? | Exchange reaction present ? |
|---|---|---|
| gln_L | ✓ | ✓ |
| glyald | no | no |
| 3hanthrn | no | no |
| citr_L | no | no |

## C) Reconstruction of transport reactions based on literature review

| Model met ID | Metabolite name | Literature Comments (Example) | Entrez Gene IDs | Reaction formula | Notes | What the example shows |
|---|---|---|---|---|---|---|
| glyald | D-Glyceraldehyde | " Na(+)-coupled entry into HIT cells involving an unidentified transport system" (PMID:7998948). | ?? | glyald[e] + na1[e] <=> glyald[c] + na1[c] | Literature review revealed that the transport is sodium coupled. Hence, this reaction is added. | No GPRs can be associated with the added reaction. Hence, integration of e.g., transcriptomic data or single gene deletion will neglect this reaction. |
| 3hanthrn | 3-Hydroxyanthranilic acid | "... The first results from a study in bladder carcinoma patients during the follow-up after curation of the tumours suggest a persistent increase in the urinary concentration of free 3-hydroxyanthranilic acid" (PMID:1244078). | ?? | 3hanthrn[c] <=> 3hanthrn[e] | Add diffusion reaction based on physiological evidence. | Additional physiological evidence are collected for the transport of the metabolite into biofluids. |
| citr_L | L-Citrulline | | ?? | citr_L[c] <=> citr_L[e] | Add diffusion reaction based on physiological evidence. | If literature review yields no clues on the transport mechanism, diffusion reactions are added. |

Figure 2

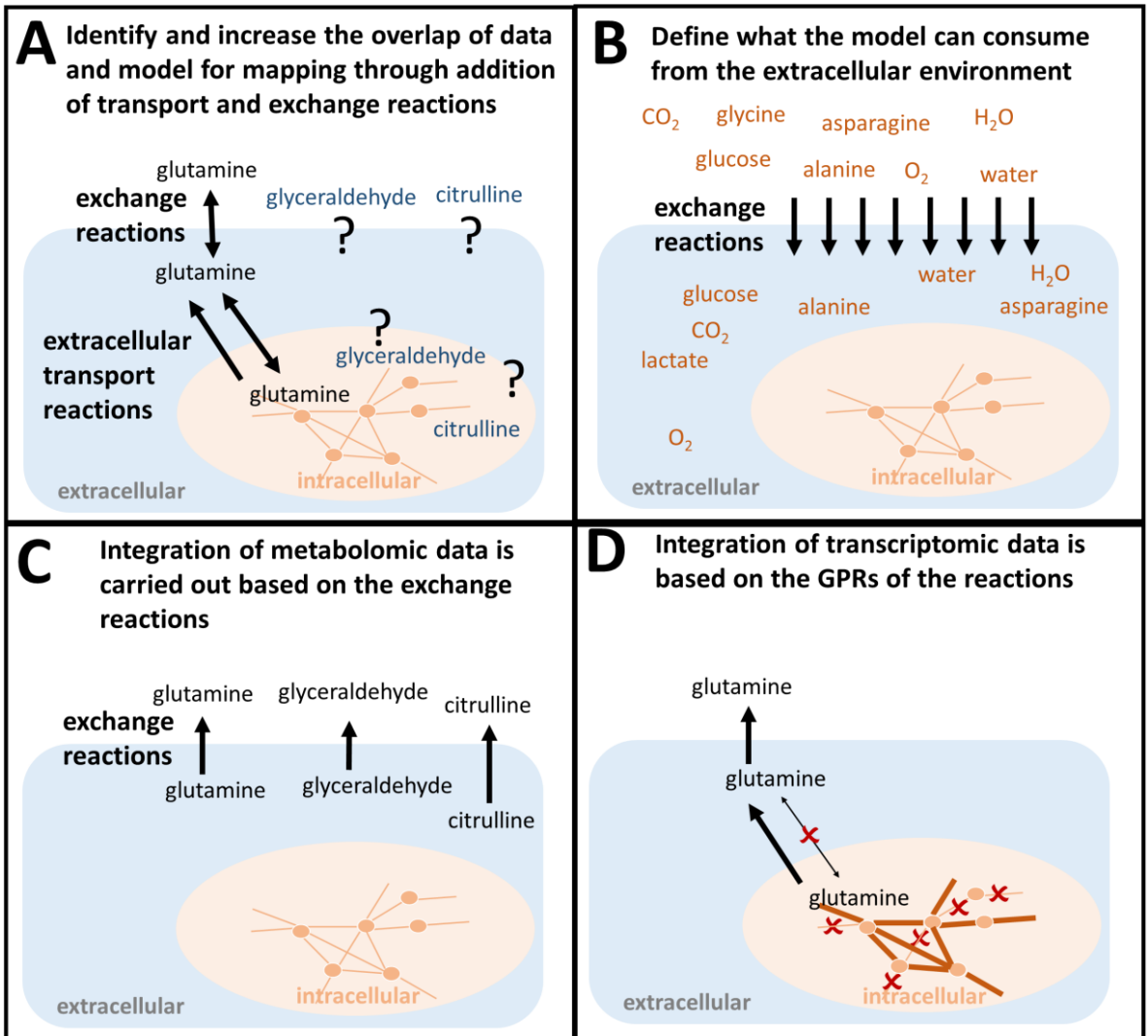

Figure 3

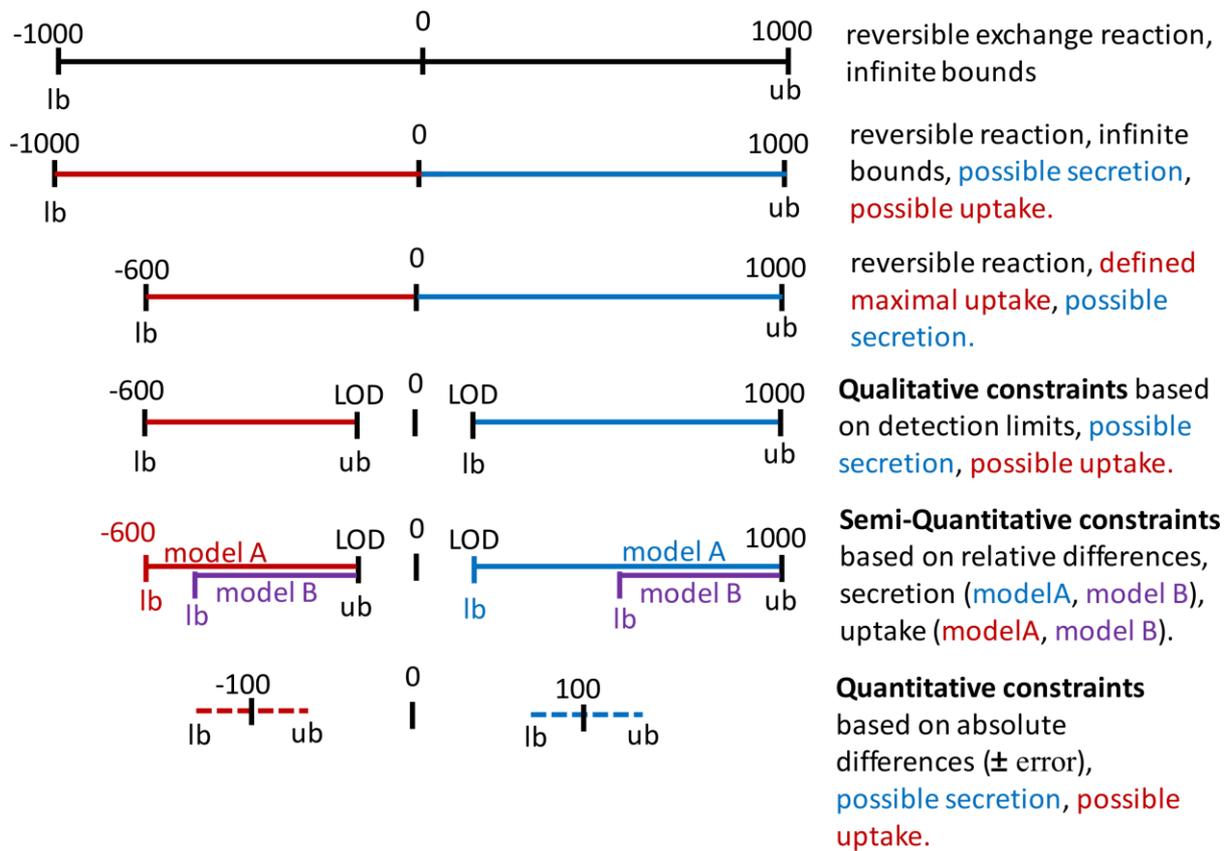

Figure 4

| A | Medium | Cell line A | Cell line B |
|---|---|---|---|
| Measured intensity at time 1 | 76555640 | 65009058 | 61697085 |
| Measured intensity at time 2 | 71459886 | 24330565 | 34981419 |
| Relative change of measured intensity: 2/1 | 0.9334 | 0.3743 | 0.5670 |
| slope (relative change: cell line/ Medium) | | 0.4010 | 0.6074 |
| slope ratio (slope: A/B) | | 0.6602 | |
| lower bound (lb) | | -18.5167 | -12.2246 |

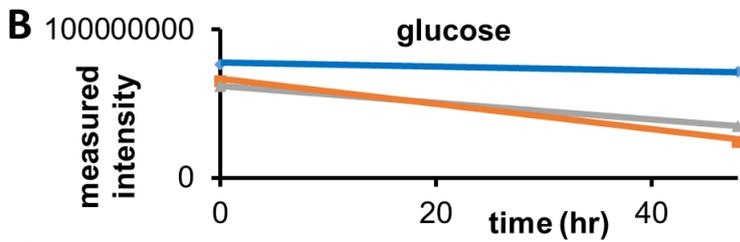

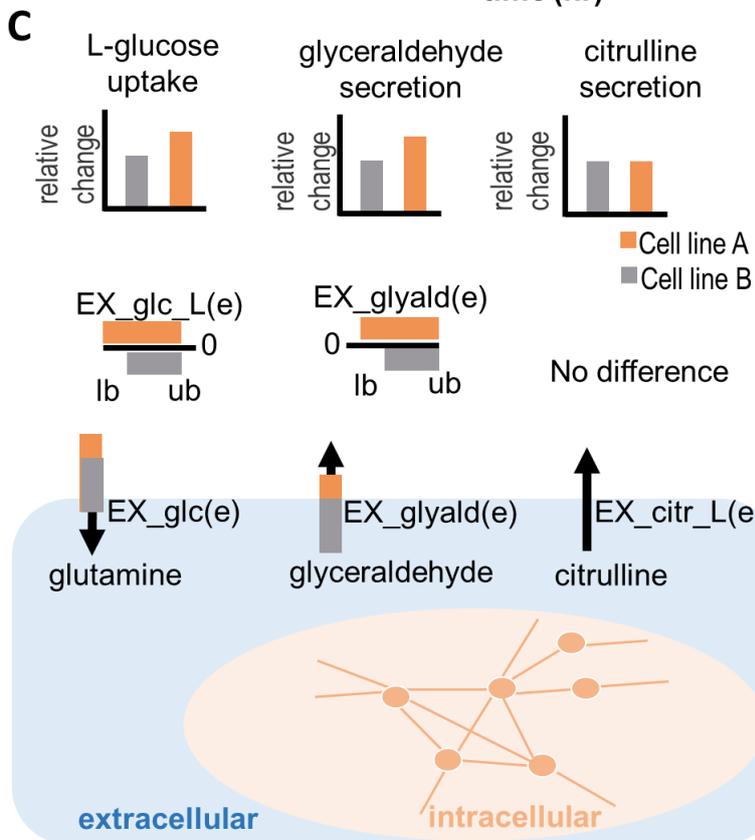

Figure 5

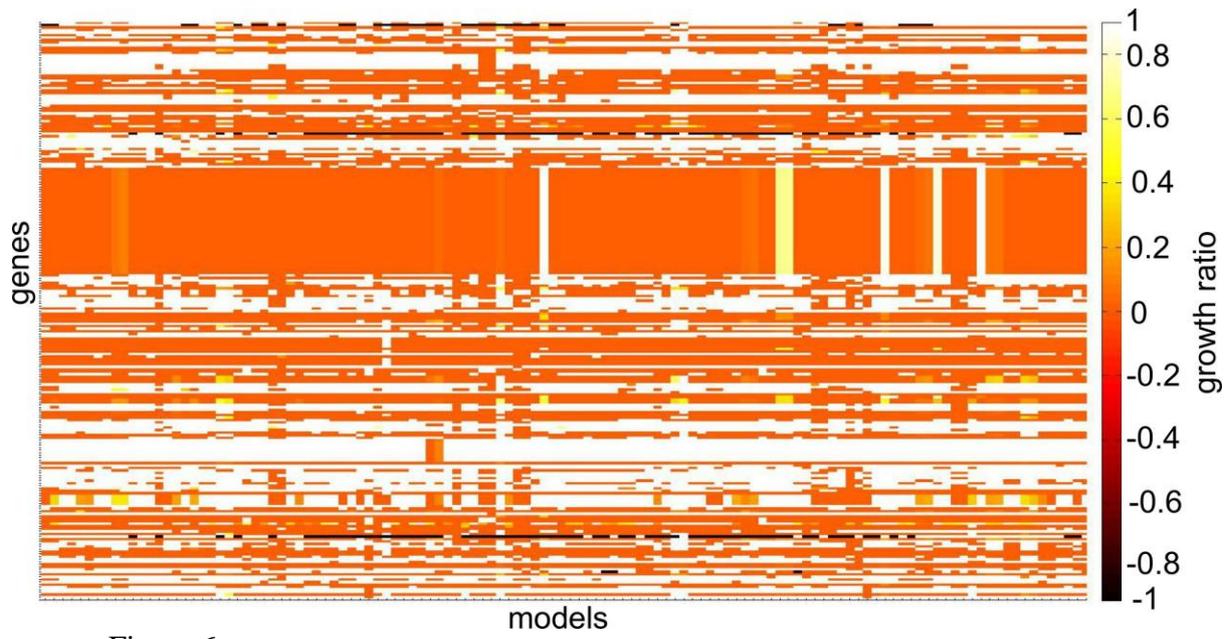

Figure 6

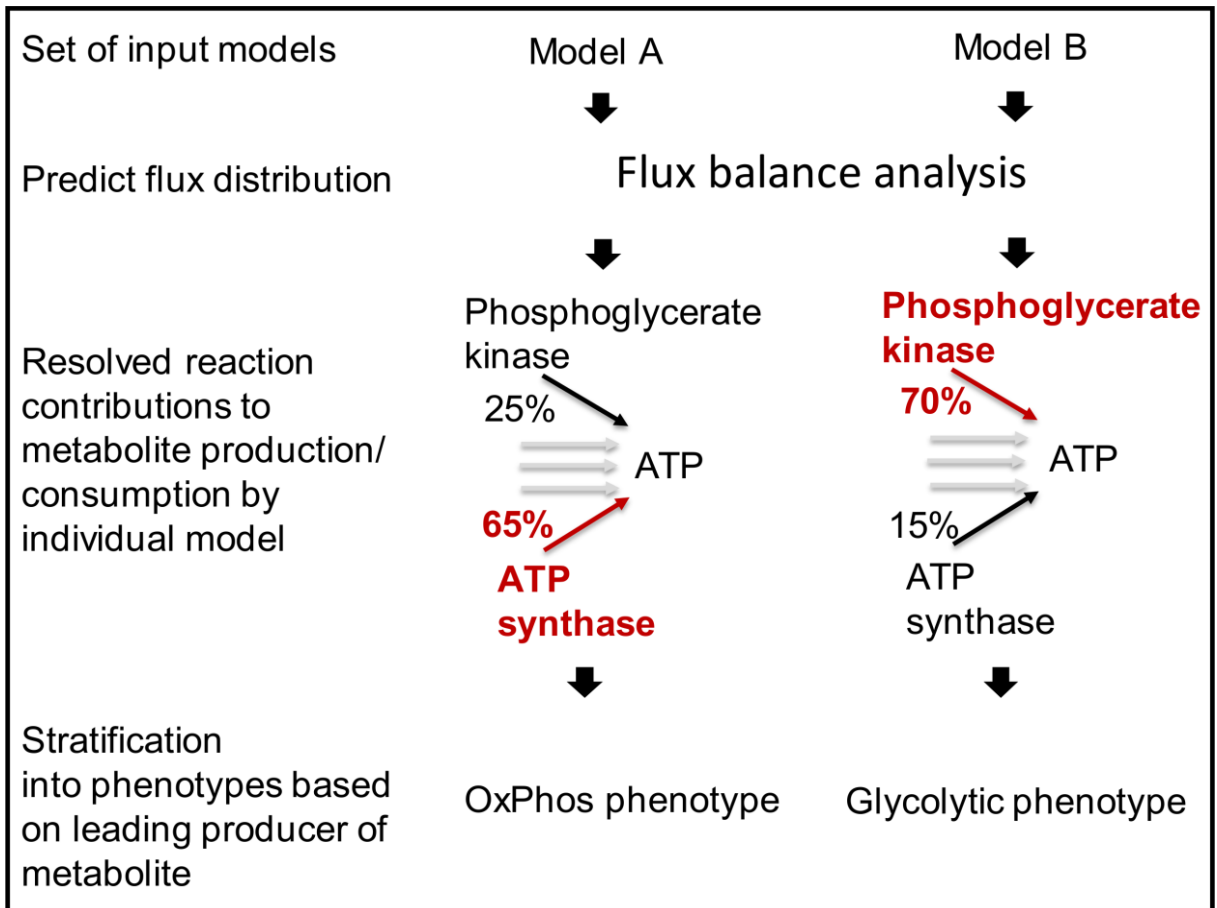

Figure 7

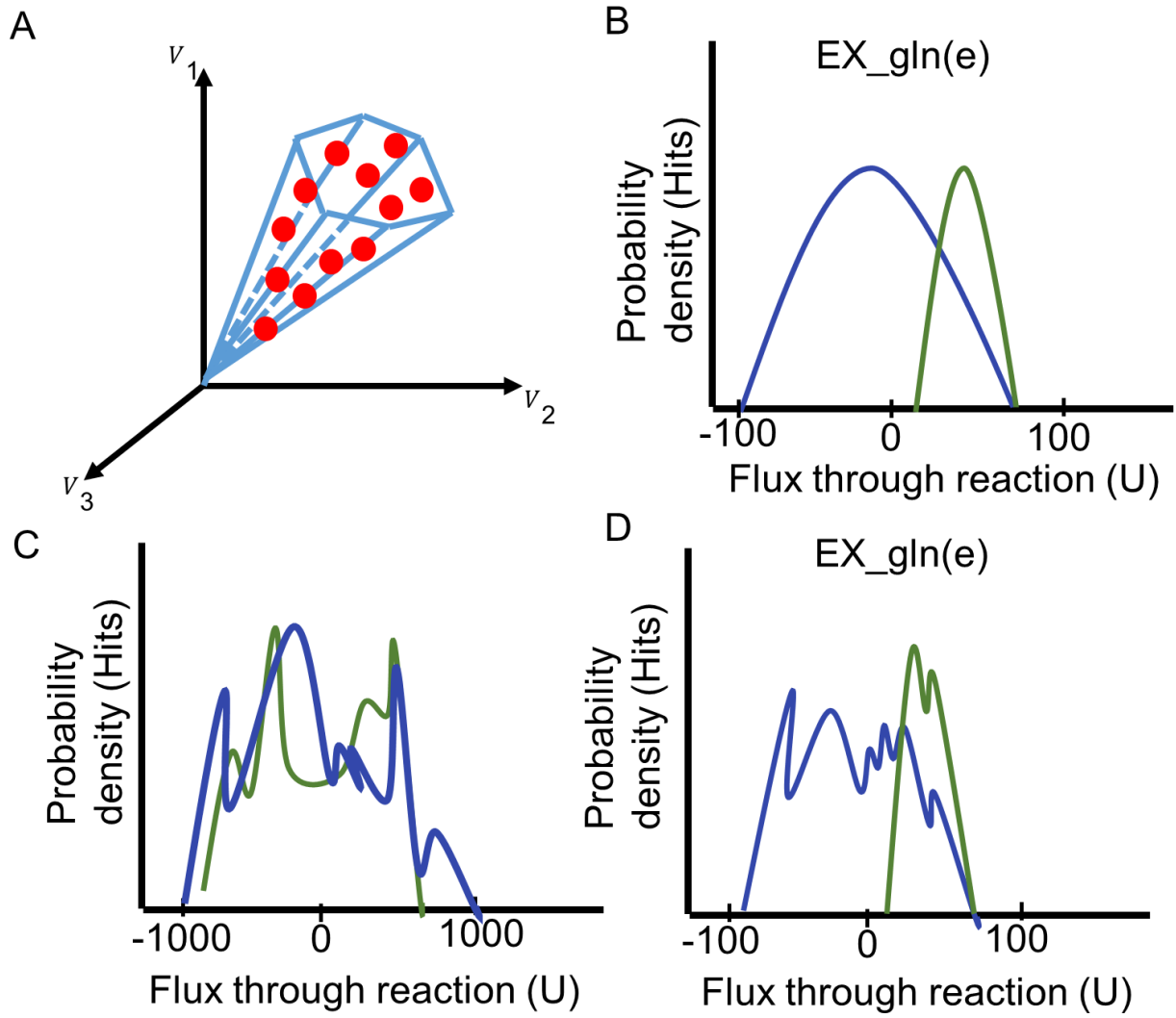

Figure 8